\documentclass[final,5p,times,twocolumn]{elsarticle}

\usepackage{hyperref}
\usepackage{amssymb}
\usepackage{subfigure}
\usepackage{epsfig}
\usepackage{epstopdf}
\usepackage{algorithm}
\usepackage{xcolor}
\usepackage{graphicx}
\usepackage{amsmath}
\usepackage{algorithm}
\usepackage[noend]{algpseudocode}

\biboptions{authoryear}

\makeatletter
\def\BState{\State\hskip-\ALG@thistlm}
\makeatother

\begin{document}

\begin{frontmatter}

\title{Mobility cost and degenerated diffusion in kinesis models}

\author[LeicMath,NN]{A.N. Gorban\corref{cor1}}
\ead{a.n.gorban@le.ac.uk}
\author[LeicMath]{N. \c{C}abuko\v{g}lu}
\ead{nc243@le.ac.uk}

\address[LeicMath]{Department of Mathematics, University of Leicester, Leicester, LE1 7RH, UK}
\address[NN]{Lobachevsky University, Nizhni Novgorod, Russia}
\cortext[cor1]{Corresponding author}


\begin{abstract}
A new critical effect is predicted in population dispersal. It is based on the fact that a trade-off between the advantages of mobility and the cost of mobility breaks with a significant deterioration in living conditions. The recently developed model of purposeful kinesis (Gorban \& \c{C}abuko\v{g}lu, Ecological Complexity 33, 2018) is based on the ``let well enough alone" idea: mobility decreases for high reproduction coefficient  and, therefore, animals stay longer in  good conditions and leave quicker bad conditions. Mobility has a cost, which should be measured in the changes of the reproduction coefficient. Introduction of the cost of mobility into the reproduction coefficient leads to an equation for mobility. It can be solved in a closed form using Lambert $W$-function. 
Surprisingly, the ``let well enough alone" models with the simple linear cost of mobility have an intrinsic phase transition: when conditions worsen then the mobility increases up to some critical value of the reproduction coefficient. For worse conditions, there is no solution for mobility.  We interpret this critical effect as the complete loss of mobility that is degeneration of diffusion.  Qualitatively, this means that mobility increases with worsening of conditions up to some limit, and after that, mobility is nullified.
\end{abstract}

\begin{keyword}
kinesis \sep diffusion \sep phase transition \sep critical effect  \sep population  \sep Allee effect

\end{keyword}

\end{frontmatter}


\section{Introduction}

The study of two basic mobility mechanisms, {\em kinesis} and {\em taxis}, is concerned with responses of organisms motions to environmental stimuli: if such a response has the form of directed orientation reaction then we call it taxis, and the change in the form of undirected locomotion is called kinesis. These `innocent' definitions cause many problems and intensive conceptual discussion \citep{Dunn1990}. One of the problems is: how to select the proper frame for discussion of the directed motion and separate the directed motion from the motion of the media. If the frame is selected unambigously then in the PDE (partial differential equations) approach to modelling taxis corresponds to change of {\em  advection} terms, whereas kinesis is modeled by the changes of the {\em mobility coefficient}.

The notion of `mobility coefficient' (or simply `mobility' for brevity)  was developed by  \citet{EinsteinBrownian} (for historical review we refer to \citet{Philibert2005}). It is summarised by the Teorell formula \citep{Teorell1935,Gorbanetal2011}

{\bf Flux = mobility$\times$concentration$\times$specific force.}

Teorell studied electrochemical transport and measured specific force as force per `gram-ion'. For ecological models \citep{Lewisetal} concentration of animals $u$  is used. The `diffusion force' is $-\nabla(\ln u)=-\frac{\nabla u}{u}$ (the `physical' coefficient $RT$ is omitted).

The most important part of Einstein's mobility theory is that the mobility coefficient is included in  the responses to {\em all forces}. For the applications of the mobility approach to dispersal of animals this means that intensity of kinesis and taxis should be connected: for example, decrease of mobility means that both taxis and kinesis decrease proportionally.

The kinesis strategy controlled by the locally and instantly evaluated well-being can be described in simple words: Animals stay longer in good conditions and leave  more quickly bad conditions. If the well-being is measured by the instant and local reproduction coefficient then the diffusion model of kinesis gives for mobility $\mu_i$ of $i$th species \citep{GorCabuk2018}:
\begin{equation}\label{ExpMob}
\mu_i=D_{0i}{\rm e} ^{-\alpha_i r_i(u_1,\ldots,u_k,s)}
\end{equation}
The corresponding diffusion equation is
\begin{equation}\label{KinesisModel}
 \partial_t u_ i ( x,t)
   =   {\rm div}[\mu_i(u_1,\ldots,u_k,s) \nabla u_i ]+r_i (u_1,\ldots,u_k,s) u_i,
   \end{equation}
   where:
\begin{itemize}
\item[]$k$ is the number of species (in this paper, we discuss  mainly the simple case $k=1$),
\item[]$u_i$ is the population density of  $i$th species,
\item[]    $s$ represents the abiotic characteristics of the living conditions (can be multidimensional),
\item[]  $r_i$ is the reproduction coefficient of $i$th species, which depends on all $u_i$ and on  $s$,
\item[]  $D_{0i}>0$ is the equilibrium mobility of $i$th species (`equilibrium' means here that it is defined for $r_i=0$),
\item[]   The coefficient $\alpha_i>0$  characterises dependence of the mobility coefficient of $i$th species on the corresponding reproduction coefficient.
\end{itemize}

This model aimed to describe the `purposeful' kinesis \citep{GorCabuk2018} that helps animals to increase there fitness when the conditions are bad (for low reproduction coefficient mobility increases and the possibility to find better conditions may increase) and not to decrease fitness when conditions are good enough (for high values of reproduction coefficient mobility decreases). The instant quality of conditions is measured by the local and instant reproduction coefficient.

\citet{GorCabuk2018} demonstrated on a series of benchmarks for models (\ref{KinesisModel}) with mobilities (\ref{ExpMob}) that:
\begin{itemize}
  \item If the food exists in low-level uniform background concentration and in rare (both in space and time) sporadic patches then  purposeful kinesis (\ref{KinesisModel}) allows animals to utilise the food patches more intensively;
  \item If there are fluctuations in space and time of the food density $s$ then purposeful kinesis  (\ref{KinesisModel})   allows animals to utilize these fluctuations more efficiently.
  \item If the presence of the Allee effect the kinesis strategy formalised by (\ref{KinesisModel})   may delay the spreading of population
  \item  The ``Let well enough alone" strategy (\ref{ExpMob}), (\ref{KinesisModel}) can prevent the effects of extinction caused by too fast diffusion  and decrease the effect of harmful diffusion described by \citet{Cosner2014}.
  \end{itemize}

The `let well enough alone' assumption (\ref{ExpMob}), (\ref{KinesisModel}) provides the mechanism for staying in a good location because mobility decreases exponentially with the reproduction coefficient. High mobility for unfavorable conditions allows animals to find new places with better conditions and seems to be beneficial.  Nevertheless, it is plausible that increase of mobility in adverse conditions requires additional resources and, therefore, there exists a negative feedback from higher mobility to the value of the reproduction coefficient. This is the `cost of mobility.' In the next section we introduce the cost of mobility and analyse the correspondent modification in the mobility function.

\section{Cost of mobility}

The `cost of mobility' has been introduced and analysed for  various research purposes. It is a well known notion in applied economic theory \citet{Tiebout1956}. The `psychic cost of mobility' and it influence on the human choice of occupations has also been discussed \citep{Schwartz1973}.  Analysis of   evolution of social traits in communities of animals demonstrated that the cost of mobility has a major impact on the origin of altruism because it determines whether and how quickly selfishness is overcome \citep{LeGalliardetal2004}. Different costs of mobility on  land and in the sea is considered as an important reason of higher diversity  on land that in the sea \citep{Vermeij2010}.  It was mentioned that the eenrgy cost of mobility may lead to surprising evolutionary dynamics \citep{Adamson2012}.

The optimality paradigm of movement is the  key part of the modern movement ecology paradigm \citep{Nathan2008}. Movement can help animals to find better conditions for foraging, thermoregulation, predator escape, shelter seeking, and reproduction. That is, movement can result in increase of  the Darwinian fitness (the average in time and generations reproduction coefficient). At the same time, movement requires spending of resources: time, energy, etc. This means that movement can decrease fecundity. The  trade-off between fecundity loss and possible improvement of conditions   is the central problem of evolutionary ecology of dispersal. In general, it is hardly known if and how mobility transfers to fitness costs. The fecundity costs of mobility in some insects was measured in field experiment  (in non-migratory, wing-monomorphic grasshopper, {\em Stenobothrus lineatus}) \citep{Samietz2012}. For some other insects (the Glanville fritillary butterfly {\em Melitaea cinxia}) the fecundity cost of mobility was not found \citep{Hanski2006}. These results challenge the hypothesis about dispersal--fecundity trade-off. A physiological trade-off between high metabolic performance  reduced maximal life span was suggested instead. Another source of the  cost of mobility may be increase of the rate of mortality due to the losses on the fly.

From the formal point of view, all types of `mobility cost' can be summarised in the negative feedback from the mobility to the reproduction coefficient: increase of mobility decreases the reproduction coefficient directly. On the other hand,  the change of conditions can increase the fitness.  Form this point of view, there is trade-off between the direct loss of fitness due to mobility and probable increase of fitness due to condition change.

In our previous model (\ref{ExpMob}), (\ref{KinesisModel})  the trade-off between the cost of mobility and the possible benefits from mobility was not accounted  \citep{GorCabuk2018}. Let us introduce here the cost of mobility as a negative linear feedback of the mobility $\mu$ on the reproduction coefficient $r$:
\begin{equation}\label{cost}
r=r_0-C\mu,
\end{equation}
where $r_0$ depends on the population densities and abiotic environment, $C$ is the cost coefficient and $C\mu$ is the cost of mobility.

According to `let well enough alone' assumptions (\ref{ExpMob}),
$\mu=D_0 \exp(-\alpha r)$. Let us introduce $\mu_0=D_0 \exp(-\alpha r_0)$, that is the mobility  (\ref{ExpMob}) for the system with the reproduction coefficient $r_0$ instead of the coefficient $r$ (\ref{cost}) with the cost of diffusion. Obviously, $\mu_0\geq \mu$ and $\mu/\mu_0=\exp(-C\mu)$.

Simple algebra gives:

$$-\alpha C\mu_0=\alpha(r-r_0)\exp(\alpha (r-r_0)).$$

\begin{figure}
\centering
\includegraphics[width=0.8\columnwidth]{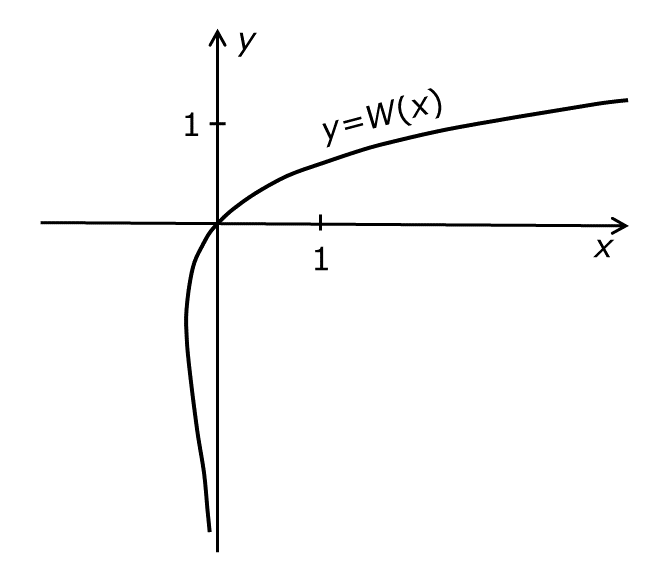}
\caption{The Lambert function $y=W(x)$ is defined for $x\geq -1/{\rm e}$. For negative $x$, the upper branch of $W$ is used, the so-called $W_0$, which is real-analytic on $(-1/{\rm e}, \infty)$. \label{Lambert}}
\end{figure}

\begin{figure}
\centering
\includegraphics[width=0.95\columnwidth]{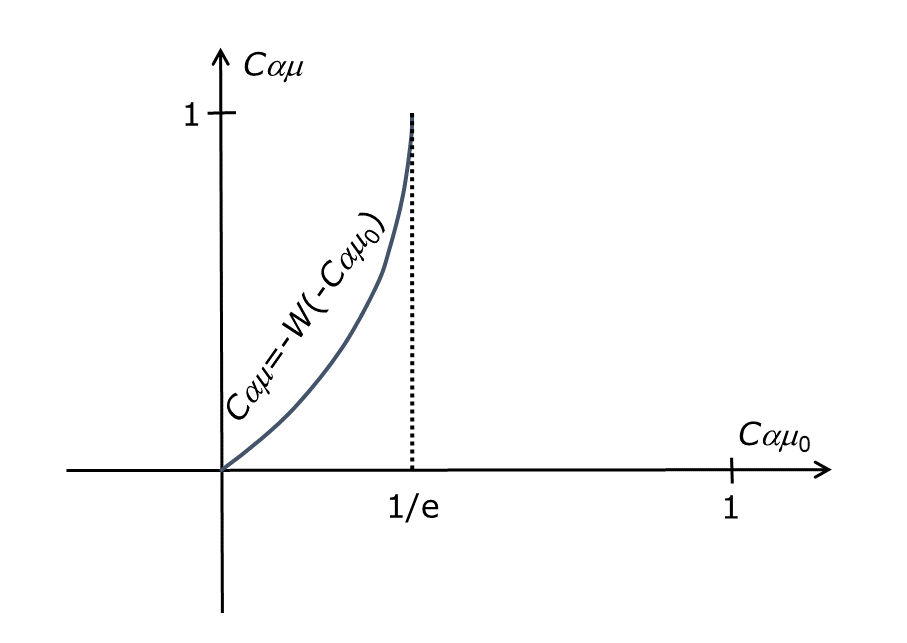}
\caption{The universal dependence of the dimensionless variable $\alpha C \mu$ on the dimensionless variable $\alpha C \mu_0$ for all models of the form (\ref{ExpMob}), (\ref{KinesisModel}) with the cost of mobility (\ref{cost}). When $C\alpha \mu_0$ exceeds $\frac{1}{\rm e}$ then the equation for mobility $\mu$ has no solution (suggested $\mu=0$).
 \label{Bifurc}}
\end{figure}

Therefore,
\begin{equation}\label{mobilityLambert}
\mu=-\frac{W(-\alpha C\mu_0)}{\alpha C},
\end{equation}
where $W$ is the Lambert $W$-function \citep{CorlessetallLambert1996}.  The Lambert $W$-function is  the inverse function to $x\exp(x)$, Fig.~\ref{Lambert}.
Function $W(x)$ is defined for $x>-1/{\rm e}$. Therefore, the mobility $\mu$ (\ref{mobilityLambert}) exists for
\begin{equation}\label{LimitValue}
\alpha C\mu_0\leq \frac{1}{\rm e}.
\end{equation}
The argument of the function $W$ in (\ref{mobilityLambert}) belongs to the interval $ [-1/{\rm e},0)$. The dependence of the dimensionless variable $\alpha C \mu$ on the dimensionless variable $\alpha C \mu_0$ (Fig.~\ref{Bifurc}) is universal for all models of the form (\ref{ExpMob}), (\ref{KinesisModel}) with the cost of mobility (\ref{cost}).

The universal limit (\ref{LimitValue}) can be represented in terms of the reproduction coefficient: the mobility formula (\ref{mobilityLambert}) is valid for
$$r_0\geq \frac{1}{\alpha}(1+\ln(\alpha C D_0)).$$
For $r_0$ below this critical solution, the equation for mobility loses solution. This is a {\em  critical transition} \citep{ShefferEtAl2012}: a `critical thresholds' is found,  where the behavior of the systems is changing  abruptly.

Definition of mobility for $\alpha C \mu_0> \frac{1}{\rm e}$ requires additional assumptions beyond (\ref{ExpMob}), (\ref{KinesisModel}), and (\ref{cost}). We have no sufficient reasons now for the definite choice. The simplest assumption is:
\begin{equation}\label{zeroextension}
\mu=0 \mbox{  for  } \alpha C\mu_0> \frac{1}{\rm e}.
\end{equation}
This collapse to zero has some biological reasons: if the further increase of mobility leads to catastrophic decrease of the reproduction coefficient (because the cost of mobility) then the reasonable strategy is to stop the dispersal at all.

\section{Equations of population dynamics with kinesis and mobility cost}

Consider an ODE model for space-uniform populations in uniform conditions:
\begin{equation}\label{ODEMOdel}
\frac{d u_i}{dt}=r_{0i}(u_1,\ldots,u_k,s)u_i
\end{equation}
(it should be supplemented by dynamic equation for abiotic components $s$).
The correspondent reaction-diffusion equations with kinesis and the mobility cost have the following form. Three additional positive  coefficients are needed for each species: $\alpha_i$, $D_{0i}$, and $C_i$. The equations are:
\begin{equation}\label{reaction-diffusion}
\begin{split}
&\partial_t u_ i ( x,t)
   =   {\rm div} (\mu_i\nabla u_i )+r_i u_i,\\
&r_i=r_{0i}-C\mu_i, \\
& \mu_i=\left\{\begin{array}{cl}-\frac{W(-\alpha_i C_i \mu_{0i})}{\alpha C_i} &\mbox{  if  } \alpha C\mu_0 \leq \frac{1}{\rm e}; \\
0 & \mbox{  if  } \alpha C\mu_0> \frac{1}{\rm e},
\end{array} \right. \\
&\mu_{0i}=D_{0i}\exp(-\alpha r_{0i}).
\end{split}
\end{equation}

\begin{figure}
\centering
a)\includegraphics[width=0.8\columnwidth]{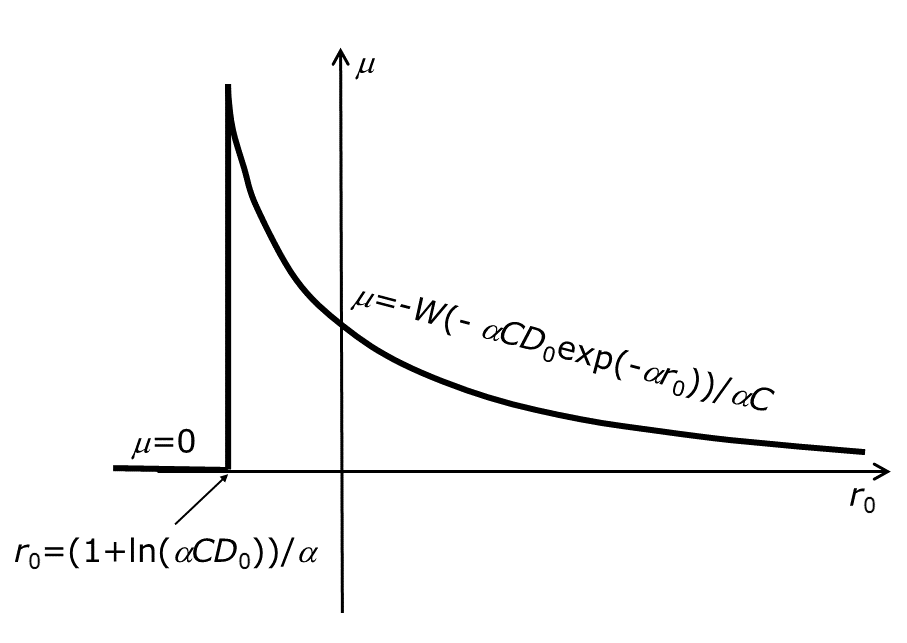}\\
b) \includegraphics[width=0.8\columnwidth]{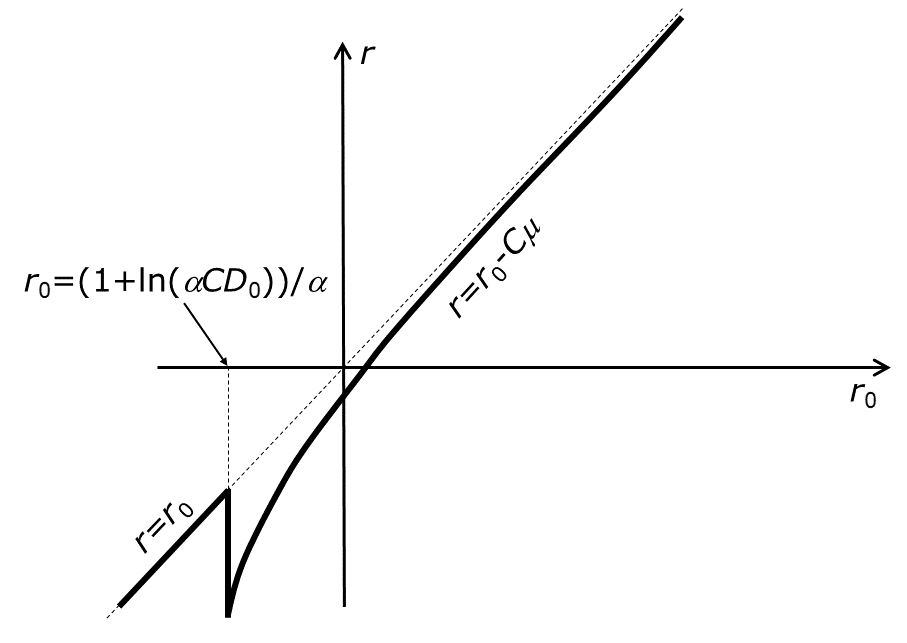}
\caption{Typical dependences of the mobility $\mu$ (a) and the modified reproduction coefficient $r$ (b) on the  unmodified reproduction coefficient $ r_0$.
 \label{MuR}}
\end{figure}

Dependence of the mobility $\mu$ on the initial reproduction coefficient $r_0$ is schematically represented in Fig.~\ref{MuR}. If $r_0$ decreases below the critical value then the mobility nullifies. This means that diffusion degenerates. Nullifying of mobility leads to increase of the reproduction coefficient $r$ because the mobility cost vanishes (see Fig.~\ref{MuR}).

Degenerating diffusion attracted much attention in the theory of porous media \citep{Vazquez2007}. The `porous media equation'  is
$$u_t=\Delta  u^m,$$
where $\Delta$ is the Laplace operator, $m>1$.

Diffusion coefficient vanishes smoothly when $u$ tends to zero.  Barenblatt \citet{Barenblatt1952} found his famous now analytic automodel solutions for equations of diffusion in porous media, and these solutions were used for modelling of nuclear bomb explosion.
Existence and regularity properties were studied in a series of works in 1960s--1970s
\citep{Aronson1969}. In 1970s, the equation of diffusion in porous media was introduced in ecological modelling \citep{GurtinMaccamy1979}. This equation predicts a finite speed of spreading of a population, which is initially confined to a bounded region.
This property is in strong contrast with the well-known properties of the classical diffusion equation, the infinite speed of propagation.

The divergent form of the porous media equation with power diffusion coefficient is
$$u_t={\rm div}(u^{\delta} \nabla u), \;\; \delta=m-1>0.$$
Exact solutions for propagation of fronts for equation
$$u_t={\rm div}(u^{\delta}\nabla u)+u^p-u^k ,$$
 were analysed for $k>p$ by \cite{Petrovskii2006}.

Equations with non-linear diffusion coefficient, which  degenerates when $u \to 0$ and goes to $\infty$ when $u \to 1$ was proposed   for modelling of  the formation and growth of bacterial biofilms \citep{Eberl2001}:
$$u_t={\rm div}(D(u)\nabla u)+ku,$$
where $ 	D(u)=\delta\frac{u^a}{(1-u)^b}$, $ a,b \geq 1 \gg \delta> 0$.
A finite difference scheme for this equation was developed and numerical experiments were provided by \citet{Eberl2007}.

Discontinuity in dependence $\mu(r_0)$ (Fig.~\ref{MuR}) causes an important property of sufficiently regular solutions: the normal derivative of $u$ nullifies  on the boundary of the areas of degenerations.  Equations (\ref{reaction-diffusion}) with non-linear mobility coefficient $\mu$  form a new family of degenerate reaction-diffusion equations. The degenerate diffusion equations  appears in many physical applications and in
geometry (Ricci flow on surfaces, for example). The typical questions are:
\begin{itemize}
\item Short and long time existence and regularity;
\item Dynamics of boundaries of degenerated areas;
\item Formation of singularities;
\item Existence through the singularities.
\end{itemize}
We believe that the detailed analysis of these equations will produce many interesting questions and unexpected answers.

\begin{figure}
\centering
\includegraphics[width=0.8\columnwidth]{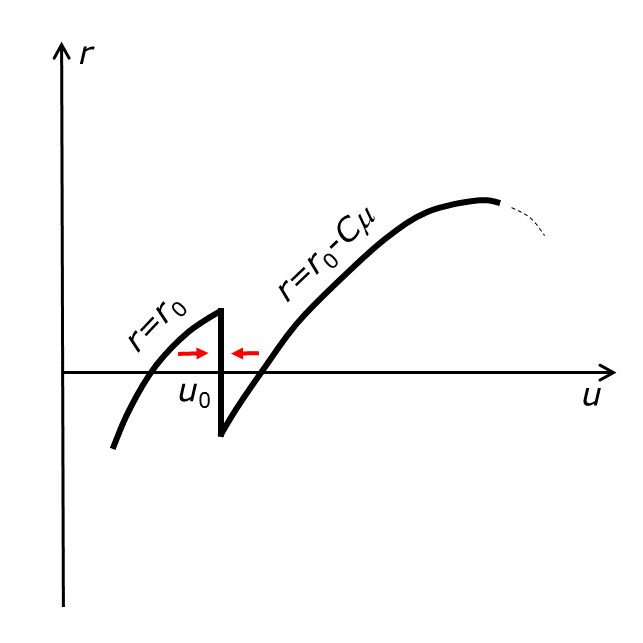}
\caption{Dependence of the reproduction coefficient $r$  on the population density for a system with the Alley effect. A special case is presented when $r(u_0+0)<0$ and $r(u_0-0)>0$. In this situation, the population dynamics $\dot{u}=r(u)u$ stabilises $u$ at the critical value (red arrows indicate the directions of changes).
 \label{RUcatasr}}
\end{figure}

Consider a system with the Alley effect to demonstrate an example of non-trivial problem and interesting effect. For such a system the reproduction coefficient $r_0(u)$ grows with $u$ on some interval. Let the critical effect appear on this interval, at point $u=u_0$ and, in addition,  $r(u_0+0)<0$ and $r(u_0-0)>0$ (Fig.~\ref{RUcatasr}). Under these conditions, the population dynamics $\dot{u}=r(u)u$ stabilises $u$ at the critical value $u=u_0$.

The solution of nonlinear equation $u_t={\rm div}(\mu(u) \nabla u) + r(u) u$ should be rigorously defined near the singularities. Instead of general definitions we apply the regularisation and transform the equation in a vicinity of the singularity into a singular perturbed system with fast relaxation.  Consider an $\varepsilon$ vicinity of $u_0$ and the equation  for $v=u-u_0$ (assume that $0<v<\varepsilon$):
\begin{equation}
\begin{split}
v_t=&{\rm div}\left(\mu(u_0+0)\frac{v}{\varepsilon} \nabla v\right)\\
& + \left(r(u_0-0)+\frac{v}{\varepsilon}(r(u_0+0)-r(u_0-0))\right)(v+u_0).
\end{split}
\end{equation}
Here, $\mu=v{\mu(u_0+0)}/{\varepsilon}$, $r=r(u_0-0)-{v}(r(u_0-0)-r(u_0+0))/{\varepsilon}$.

Solution of this equation stabilises at $v=\varepsilon r(u_0-0)/(r(u_0-0)-r(u_0+0))$. At this state, $r=0$ and
$$\mu=\frac{\mu(u_0+0)r(u_0-0)}{r(u_0-0)-r(u_0+0)}.$$
Therefore, there appear areas with (almost) critical value of the population density $u\approx u_0$ and effective reproduction coefficient $r\approx 0$. This appearance of areas with constant critical density and equilibrium (zero) reproduction coefficient resembles the growth of  biofilm \citep{Eberl2001}.

\section{Generalizations}

The observed effect is not a special property of the Lambert function and is robust. Consider equations (\ref{KinesisModel}) with mobility function
\begin{equation}\label{ConvMob}
\mu_i=D_{0i}h(-\alpha_i r_i(u_1,\ldots,u_k,s)),
\end{equation}
where $h(z)>0$ is a monotonically growing, convex, and  twice differentiable function on real axis, $h''(z)>0$ and $h'(z)\to \infty$ when $z \to \infty$ (this $h(x)$ substitutes exponent in (\ref{ExpMob})).

Using the same linear cost of mobility $C\mu$ (\ref{cost}) we get
\begin{equation} \label{mobilityEq}
r=r_0-CD_0h(-\alpha r)
\end{equation}
 or
\begin{equation}\label{critical}
h(y)=\frac{y}{CD_0\alpha}+\frac{r_0}{CD_0},
\end{equation}
where $y=-\alpha r$.
There exists a unique solution $y_c$ of the equation $$h'(y)=\frac{1}{CD_0\alpha}.$$ Therefore, for solutions of equation (\ref{critical}) we get:
\begin{itemize}
\item If $r_0>CD_0h(y_c)-({y_c}/{\alpha})$ then (\ref{critical}) has two solutions;
\item If $r_0=CD_0h(y_c)-({y_c}/{\alpha})$ then  (\ref{critical}) has one solution $y=y_c$;
\item If $r_0<CD_0h(y_c)-({y_c}/{\alpha})$ then  (\ref{critical}) has no solutions.
\end{itemize}

Qualitatively, the situation is the same as for the exponent: there exists a critical value of the reproduction coefficient $r_0$ and when it decreases below this critical value,  then the equation for mobility has no solution. The explicit solution with Lambert function allowed us a bit more: we found  the universal explicit dependence between dimensionless quantities $y=C\alpha \mu$ and $v=C \alpha \mu_0$,  $y=-W(-v)$ (Fig.~\ref{Bifurc}), which does not change with parameters.

\begin{figure}
\centering
\includegraphics[width=0.8\columnwidth]{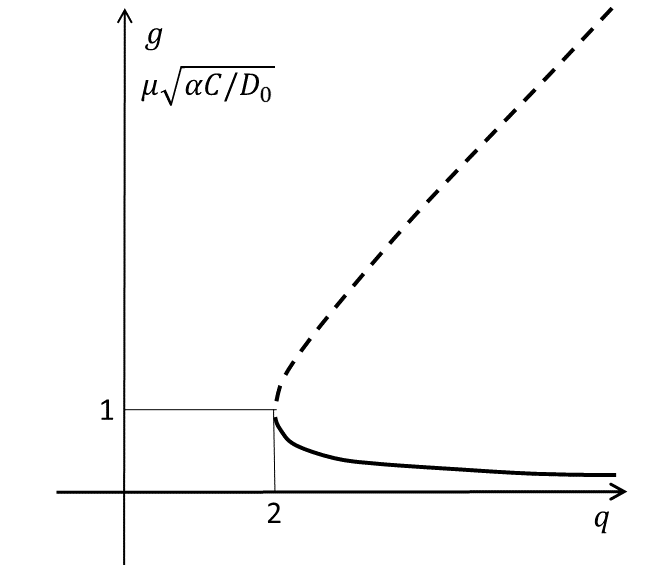}
\caption{The universal dependence of the dimensionless variables $g=(\alpha r+1)/\sqrt{CD_0}$ (upper branch, dashed line) and $\mu \sqrt{{\alpha C}/{D_0}}$ (bottom branch, solid line) on the dimensionless variable $q=(\alpha r_0+1)/\sqrt{CD_0}$ for models of the form (\ref{ExpMob}), (\ref{ConvMob}) with the cost of mobility (\ref{mobilityEq}) and $h(z)=1/(1-z)$. The equation for mobility  has no solution (suggested $\mu=0$) when $q<2$. \label{SQRTmu}}
\end{figure}

For simple algebraic functions $h$ (proposed by an anonymous MDPI reviewer) the universal explicit  solutions are also possible. Consider
$$h(z)=\frac{1}{1-z}.$$
This function is defined for $z<1$, is convex on this semi-axis, $h''(z)>0$, and $h'(z)\to \infty$ when $z \to 1$. Let us use this $h$ in (\ref{ConvMob}). Solution of equation (\ref{mobilityEq}) is

$$g=\frac{q}{2}+\sqrt{\frac{q^2}{4}-1},\;\; \sqrt{\frac{\alpha C}{D_0}}\mu=\frac{q}{2}-\sqrt{\frac{q^2}{4}-1}$$
where the dimensionless variables $g$ and $q$ are:
\begin{itemize}
\item[] $g=(\alpha r+1)/\sqrt{CD_0}$,
\item[] $q=(\alpha r_0+1)/\sqrt{CD_0}$.
\end{itemize}
Solution exists if $q\geq 2$ and does not exist if $q<2$ (i.e. $r_0<(2\sqrt{CD_0} -1)/\alpha$) (see Fig.~\ref{SQRTmu}).

\section{Mobility and relation between spatial and temporal correlations}

Kinesis could be beneficial for animals because it allows them to find better conditions. The probability distribution of such benefits depends on correlations of conditions in space and time. Qualitatively, if correlation in space are low for bad conditions  then it is possible to find better conditions with random movement. If correlation in time are high then the strategy `to wait' can be worse than the strategy `to move' because the probability that the situation will become better at the same place is smaller than the probability to find better conditions in random walk. In the opposite case, when the correlations in space are high, and the correlation in time are small, then it may be more beneficial to wait at the same place then to move.

The  benefits from motion should be compared to the mobility cost. Both these quantities should be measured in the reproduction coefficient. The interplay between these quantities determines the optimal kinesis strategy.

Detailed analysis of the optimal mobility by the methods of the evolutionary optimality (see, for example  works by \citet{Hofbauer1998, Gorban2007}),  requires more detailed models and much more data. Dynamics of the adaptation resource of animals spent for mobility \citep{Gorban2016} and the typical spatial and temporal correlations of conditions should be taken into account.

Nevertheless, qualitative analysis of benefits from kinesis for various space and time correlations is very desirable. Let us simplify the problem and discuss discrete space (two locations) and time. The following simple example demonstrates how the `stop mobility' effect depends of the relations between the spatial correlations, the temporal correlations and the cost of mobility.

Let us start from the simple model used by \citet{GorCabuk2018} to illustrate the idea of purposeful kinesis. An animal can use one of two locations for reproduction. The environment in these locations can be in one of two states during the reproduction period, $A$ or $B$. The number of surviving descendants is $r_A$ in state $A$ and $r_B$ in state $B$.  Their further survival does not depend on this area.
Let us take $r_A>r_B$ (just for concreteness).

The animal can just evaluate the previous state of the locations where it is now but cannot predict the future state. There is no memory: it does not remember the properties of the locations where it was before. It can either select the current (somehow chosen)  location or to move to another one. It can do no more than one change of locations. The change of location decreases the reproduction coefficient by multiplication on ${\rm e}^{-C}$ (cost of mobility).

Let $S_1(t)$ and $S_2(t)$ be the states of the locations 1 and 2, correspondingly. Assume also that changes of the pairs $(S_1,S_2)$ can be descried by an ergodic Markov chain with four states $(A,A)$, $(A,B)$, $(B,A)$, and $(B,B)$. Let all the transition probabilities be symmetric with respect to the change of locations $1\leftrightarrow 2$. Four conditional probabilities are needed for analysis of mobility effects in this model:
$${\mathbf P}(S_1(t+1)=A|S_1(t)=A),  \; {\mathbf P}(S_1(t+1)=B|S_1(t)=A), $$
$${\mathbf P}(S_2(t+1)=A|S_1(t)=B),  \; \mbox{  and  } \; {\mathbf P}(S_2(t+1)=B|S_1(t)=B).$$
Assume that an animal is at time $t$ in the location with state $A$, then:
\begin{itemize}
\item if the the animal remains in the initial location  then the expected number of surviving descendants  is $${\mathbf P}(S_1(t+1)=A|S_1(t)=A) r_A +{\mathbf P}(S_1(t+1)=B|S_1(t)=A) r_B.$$
\item if the animal jumps to another location then the expected number of surviving descendants  is
\begin{equation*}
\begin{split}
{\rm e}^{-C}[&{\mathbf P}(S_2(t+1)=A|S_1(t)=A) r_A\\ &+{\mathbf P}(S_2(t+1)=B|S_1(t)=A) r_B].
\end{split}
\end{equation*}
\end{itemize}
If an animal is at time $t$ in the location with state $B$, then:
\begin{itemize}
\item if the the animal remains in the initial location  then the expected number of surviving descendants  is $${\mathbf P}(S_1(t+1)=A|S_1(t)=B) r_A +{\mathbf P}(S_1(t+1)=B|S_1(t)=B) r_B.$$
\item  if the animal jumps to another location  then the expected number of surviving descendants  is
\begin{equation*}
\begin{split}
{\rm e}^{-C}[&{\mathbf P}(S_2(t+1)=A|S_1(t)=B) r_A \\&+{\mathbf P}(S_2(t+1)=B|S_1(t)=B) r_B].
\end{split}
\end{equation*}
\end{itemize}

The choice `to stay in the current location or to jump' is determined by the selection of behaviour with the highest number of expected offspring. In the evaluation of this number the temporal correlations between $S_1(t)$ and $S_1(t+1)$, the spatio-temporal  correlations between $S_1(t)$ and $S_2(t+1)$, and the cost of mobility coefficient ${\rm e}^{-C}$ are used.

\section{Discussion}

Superlinear increase of the mobility for decrease of the reproduction coefficient in combination with linear cost of mobility  leads to the critical effect: for sufficiently bad condition the solution of equation for mobility does not exist. For some dependencies of mobility on the reproduction coefficient this critical effect can be found explicitly (for example, for the exponential dependence (\ref{ExpMob}) proposed and analysed in our previous work \citep{GorCabuk2018}).

Existence of  the critical effect is proven. The question arises: how to find mobility after the critical transition? There is no formal tool to find the answer. We suggest that after the critical threshold, the mobility nullifies. Qualitatively this means that with worsening of conditions mobility  increases up to some maximal value. If the conditions  deteriorate further, another mobility strategy is activated: do not waste resources for mobility, just wait for  conditions to change.

The exact values of the critical thresholds and the optimal dependence of mobility on the reproduction coefficient depend on the correlation of the conditions in space and time. Typical correlations  during the evolution time should be used. These correlations are unknown, and instead plausible hypotheses and identification of parameters from the data can be used.

There are several directions of further work:
\begin{itemize}
\item We expect that the described critical effect was widespread in nature, but its description required a theoretical basis. Now this basis is proposed, and existing data on animal mobility can be revised  to understand the new critical effect.
\item The new family of models requires additional theoretical (mathematical) and numerical analysis with the development of existence and uniqueness theorems, the analysis of attractors, and the development of adequate numerical methods.
\item It would be great to apply the new models for modelling of dispersal of real population.
\end{itemize}


\section*{Acknowledgement}
AG and N\c{C} were supported by the University of Leicester, UK. AG was supported by the Ministry of education and science of Russia (Project No. 14.Y26.31.0022)

\end{document}